\documentclass[conference]{IEEEtran}

\usepackage[pdftex]{graphicx}
\usepackage{enumitem}
\usepackage{float}
\usepackage[caption = false]{subfig}
\usepackage{flushend}
\usepackage[export]{adjustbox}

\begin{document}

\title{A Machine Learning Analysis of Twitter Sentiment to the Sandy Hook Shootings}

\author{\IEEEauthorblockN{Nan Wang, Blesson Varghese}
\IEEEauthorblockA{School of EEECS\\
Queen's University Belfast, UK\\
Email: \{nwang03, varghese\}@qub.ac.uk}
\and
\IEEEauthorblockN{Peter D. Donnelly}
\IEEEauthorblockA{Dalla Lana School of Public Health\\
University of Toronto, Canada\\
Email: peter.donnelly@oahpp.ca}
}

\maketitle

\begin{abstract}
Gun related violence is a complex issue and accounts for a large proportion of violent incidents. In the research reported in this paper, we set out to investigate the pro-gun and anti-gun sentiments expressed on a social media platform, namely Twitter, in response to the 2012 Sandy Hook Elementary School shooting in Connecticut, USA. Machine learning techniques are applied to classify a data corpus of over 700,000 tweets. The sentiments are captured using a public sentiment score that considers the volume of tweets as well as population. A web-based interactive tool is developed to visualise the sentiments and is available at http://www.gunsontwitter.com. The key findings from this research are: (i) There are elevated rates of both pro-gun and anti-gun sentiments on the day of the shooting. Surprisingly, the pro-gun sentiment remains high for a number of days following the event but the anti-gun sentiment quickly falls to pre-event levels. (ii) There is a different public response from each state, with the highest pro-gun sentiment not coming from those with highest gun ownership levels but rather from California, Texas and New York.

\end{abstract}

\IEEEpeerreviewmaketitle

\section{Introduction}
\label{introduction}
On the morning of $14^{th}$ December 2012 a heavily armed twenty-year-old man with \textit{``significant mental health issues''} \cite{Sedensky} and access to his mother's legally owned firearms shot his way into the locked building of the Sandy Hook Elementary School (SHES) in Connecticut, USA and in less than eleven minutes murdered 20 children and 6 adults. Before driving to the school he had shot and killed his mother in her bed and at the conclusion of the shootings he took his own life.

The world media response to this tragedy was unprecedented in its scale and much has subsequently been written about the events themselves and how future similar events could be avoided \cite{shes-1}. Predictably much attention focussed on issues of gun ownership and control with those advocating tighter gun control clashing with those intent on exercising and protecting (as they saw it) the US constitutional second amendment right to bear arms. That debate continues and, in its totality, is beyond the scope of this paper in which we largely restrict ourselves to seeking to understand and interpret pro-gun and anti-gun sentiment as expressed on a social media platform, namely Twitter in a forty day period starting seven days before the shootings. 
 
In order to achieve this we have used the application of machine learning to a sample of over 700,000 tweets made by individuals in the United States that contain one or more predetermined relevant key words. What we sought to capture was pro-gun and anti-gun sentiment and how it changed over time. For this we developed a framework to collect, pre-process and classify tweets and further visualise the sentiments. A hand curated gold standard dataset is employed in classifying the entire sample. A number of machine learning approaches are evaluated and the approaches that provide maximum accuracy over a small sample are used for classifying the entire tweet sample. We visualise this sentiment by state taking into account state population size and gun ownership levels through an application we have hosted at http://www.gunsontwitter.com.

The key observations are that there is a peak of both anti-gun and pro-gun sentiment on the day of the shooting. Anti-gun sentiment quickly falls to pre-event levels whereas pro-gun sentiment remains high for a number of days. It is interesting to note that all states do not respond in the same way; the highest gun ownership states are not the most pro-gun, but California, Texas and Florida have highest pro-gun sentiment even when their larger population size is taken into account. 

The remainder of this paper is organised as follows. 
Section \ref{methodology} presents the methodology adopted in this paper to collect, pre-process, classify, summarise and visualise the sentiment of the tweets.
Section \ref{mlapproaches} considers the machine learning approaches employed for classifying the tweets. 
Section \ref{casestudy} takes the Sandy Hook school shooting case study into account by applying the methodology on a volume of tweets and visualising pro-gun and anti-gun sentiments. 
Section \ref{discussion} presents the public health implications of the research reported in this paper.
Section \ref{conclusions} concludes this paper.

\section{Methodology}
\label{methodology}
\begin{figure}
	\centering
	\includegraphics[width=0.48\textwidth]{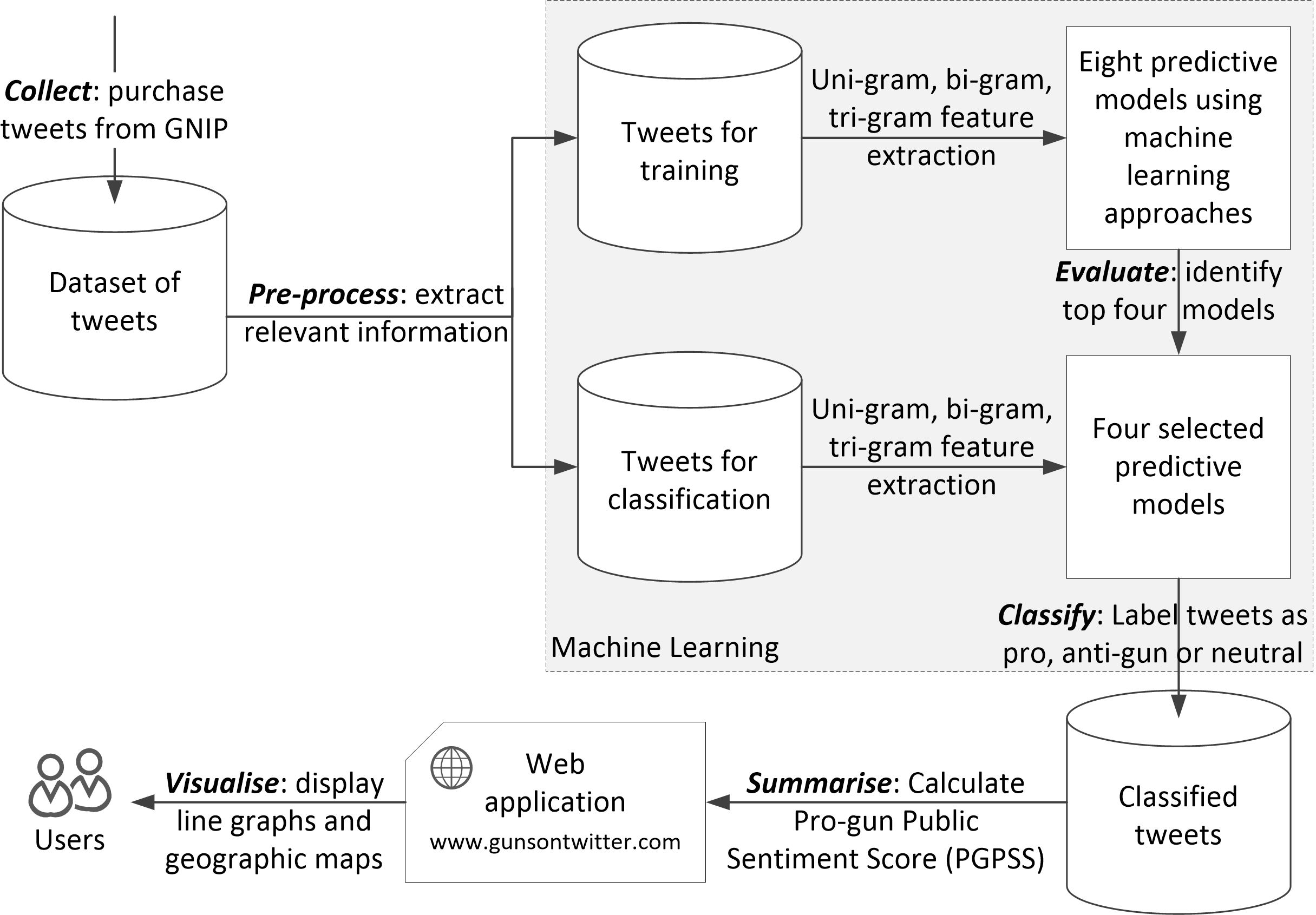}
	\caption{Methodology for analysing and visualising public sentiment}	
	\label{methodology:figure1}
\end{figure}

The methodology for analysing public sentiment incorporates machine learning and (1) collects, (2) pre-processes, (3) classifies, (4) summarises and (5) visualises data. Figure \ref{methodology:figure1} illustrates the sequence of steps in the methodology. 

Raw data from Twitter is collected and pre-processed into a trimmed data set in an appropriate format, which is further split into training and testing data sets. In the machine learning stage different algorithms, such as Support Vector Machines (SVM), Maximum Entropy (ME), Tree, Bagged Tree, Boosted Tree, Random Forest (RF), Neural Network (NN) and Na{\"i}ve Bayes (NB), are explored to build classifiers. A subset of more accurate classifiers are used for classifying the tweets. A number of functions that can be used to derive meaningful information are applied on the classified tweets. The results are summarised using multiple visualisation techniques.

\subsubsection*{Step 1 - Collect}
The raw data of tweets in JSON format is purchased from GNIP\footnote{http://gnip.com/}, a third-party authorised to resell historical tweets determined by predefined rules that takes into account filters based on the time-period, key words and geographic location. Compared to alternatives such as custom-built web crawlers or programming the Twitter Streaming API\footnote{https://dev.twitter.com/docs/streaming-apis}, this approach saved time and provided data in a consistent format, although the tweets had to be purchased.

\subsubsection*{Step 2 - Pre-process}
A file containing a list of JSON files and their remote locations was obtained from the reseller. Each JSON file is organised for a ten minute period and contains the tweets and a large amount of information regarding the tweet\footnote{http://gnip.com/sources/twitter/historical/}. The JSON files were parsed to extract the most relevant data for the research and were converted into the CSV format. All further tasks were performed using the R programming language in RStudio\footnote{http://www.rstudio.com}. This was preferred to exploit the readily available packages for data mining and natural language processing.

\subsubsection*{Step 3 - Classify}
In the machine learning stage, the sentiments of the tweets are extracted. The training data set contains 5000 tweets selected from the trimmed data while the remaining tweets are left for classification. The sentiment of each tweet is manually labelled into three classes, namely; pro-gun, anti-gun and neutral, to obtain a gold standard dataset. It is assumed that the distribution of tweets is not equal among these classes on grounds of the characteristic of Twitter - a convenient platform for users to express subjective ideas (which convey a pro-gun or anti-gun sentiment). Therefore, the numbers of tweets for three classes are adjusted as 2000 respectively for the pro-gun and anti-gun classes and 1000 for the neutral class. Intentionally giving extra weight to subjective classes is known to produce classifiers that are more sensitive to these classes \cite{weight-1}.

Using the extracted features classifiers are built using supervised training approaches which are presented in the next section. Prior to the deployment of more accurate classifiers, the effectiveness of all models are evaluated by the output of sample accuracy in a 10-fold cross validation \cite{zhang1993model}. Only top classifiers are selected and further developed based on the features extracted for the classification of the remaining tweets.

\subsubsection*{Step 4 - Summarise}
In addition to the time stamps and coordinates that the data originally contains, a series of methods for calculating a Pro-Gun Public Sentiment Scores (PGPSS) are utilised for comparison. Consider a geographic region defined by $g$ and a time frame $t$. The baseline PGPSS is denoted as 
\begin{eqnarray*}
PGPSS_{1}&=&\frac{count_{(g,t)}\;(positive\;tweets)}{count_{(g,t)}\;(negative\;tweets)}\label{methodology:equation1}
\end{eqnarray*}

Baseline PGPSS measures the number of pro-gun tweets over anti-gun tweets from a set of tweets originating from a state in a given time frame. It gives the degree of positivity in the selected tweets as an index which measures the positive tweets over negative tweets. However, $PGPSS_1$ is not the best indicator of positivity since it does not take the volume of tweets that originates from $g$. This poses a problem which is illustrated in the following example. 

Consider the country $g$ has population of 11 million, which has two states $g_1$ and $g_2$. Suppose the population of $g_1$ is 10 million and $g_2$ is 1 million, and the total number of available tweets from the states be 1 million (200,000 positive and 800,000 negative) and 10,000 (8,000 positive and 2,000 negative) respectively. Now $PGPSS_1$ for $g_1$ is 0.25 and for $g_2$ is 4 which indicates a higher positive sentiment for $g_2$ than $g_1$. The normalised $PGPSS_1$ for $g_1$ is 0.0625 and for $g_2$ is 1. However, this is not entirely accurate in that the volume of tweets that originated from $g_2$ is relatively a lot smaller than the tweets from $g_1$. Hence, a score that takes the volume of tweets that originates from $g_1$ and $g_2$ into account is useful. 

Hence, we define a PGPSS score, which is corrected for volume of tweets defined as
\begin{eqnarray*}
PGPSS_{2}&=&\frac{count_{(g,t)}\;(positive\;tweets)}{count_{(g,t)}\;(negative\;tweets)}\;* \\
& & \frac{count_{(g,t)}\;(tweets)}{count_{(t)}\;(tweets)}\label{methodology:equation2}
\end{eqnarray*}

The volume correction ratio for $g_1$ is $1,000,000 / 1,010,000 = 0.990099$ and for $g_2$ is 0.009901. Now $PGPSS_2$ for $g_1$ is 0.247525 and for $g_2$ is 0.039604, and the normalised $PGPSS_2$ for $g_1$ is 1 and for $g_2$ is 0.16.

An additional correction for population is considered by defining a different PGPSS score which is denoted as 
\begin{eqnarray*}
PGPSS_{3}&=&\frac{count_{(g,t)}\;(positive\;tweets)}{count_{(g,t)}\;(negative\;tweets)}\;*\\
& &\frac{count_{(g,t)}\;(tweets)}{count_{(t)}\;(tweets)}\;*\frac{population_{g}}{population}\label{methodology:equation3}
\end{eqnarray*}

The population correction ratio for $g_1$ is $1 million / 11 million = 0.90909$ and for $g_2$ is $1000 / 11 million = 0.000909$. The $PGPSS_3$ score for $g_1$ is 0.225023 and for $g_2$ is 0.000036, and the normalised $PGPSS_3$ for $g_1$ is 1 and for $g_2$ is 0.000016. 

All PGPSS scores are normalised between 0 and 1 for the purpose of visualisation. To interpret the PGPSS scores in the case of public sentiment towards gun use, $g_2$ is more supportive of guns than $g_1$ according to $PGPSS_1$, but after correcting the scores, as should be $g_1$ is presented as a state more supportive of guns than $g_2$. The corrections applied to the public sentiment scores provide a better overview in the context of the volume of tweets and population.

\subsubsection*{Step 5 - Visualise}
A web application is developed under the Shiny framework\footnote{http://shiny.rstudio.com} for visualising data. Motion charts, line graphs and geographic maps are generated using Google Charts API\footnote{https://developers.google.com/chart/} through the googleVis package \cite{gesmann2013package}. These techniques present results at the country and state levels as well as hourly and daily analyses.

Motion charts display a continuous bubble view of data for a combination of six variables simultaneously; for example, the x and y axes, a bubble, its size and colour and a slider all display relevant information. This technique provides an animated overview of the changes through an interface to observe variations over time. Additionally, a bar plot and a line graph are embedded in the motion chart.

The line graph presents the volume of classified tweets over the dimension of time. A control gadget is available for users to monitor the data on daily or hourly basis. Options on zooming and changing time period are embedded in the graph.

The geographic map visualises a series of PGPSS results in the form of a state-level map. A particular type of Public Sentiment Score and time frame are selected by users each time to be projected into a gradient colour on the map. This technique is useful to interpret the statistics from a geographic perspective.

\section{Machine Learning Approaches}
\label{mlapproaches}
Feature extraction and modelling are crucial steps in any machine learning approach. Various techniques on feature selection have been examined in previous research, from the baseline approach of N-gram features \cite{kouloumpis2011twitter} to the advanced linguistic analysis on POS tags \cite{Go}. Meanwhile, SVM and NB are the top two frequently used algorithms for sentiment analysis and favoured by many researchers due to their superior performance \cite{nguyen2013royal}. The evaluation of model performances compares their accuracy from the aspect of training data sizes, N-gram features and algorithms.

\subsection{Feature Extraction}
In the feature extraction stage, uni-gram, bi-gram and tri-gram are considered in the domain of N-gram features \cite{kouloumpis2011twitter}. As the most frequently used method in text classification, feature extraction using uni-grams process each word within the sentence individually. Phrases of two and three words are extracted as Bi-gram and tri-gram features for contrasts.

Hashtags and reply/mention tags within each tweet are used as additional features in that sometimes they contain strong subjective opinions. For example in the case of sentiment towards gun use, \textit{\#2ndamendment} may indicate pro-gun views on the use of guns whereas \textit{\#guncontrol} may suggest more neutral or anti-gun views. Similarly, \textit{@NRA} may be a strong indicator of emotional text, albeit vague in terms of polarity.

\begin{table*}
	\centering
	\caption{Output of sample accuracy generated through 10-fold cross validation using uni-gram feature.}
	\begin{tabular}{|c|c|c|c|c|c|c|c|c|}
	\hline
	Training Data Size & SVM & ME & Tree & Bagged Tree & Boosted Tree & RF & NN & NB \\ \hline \hline
	1000 & 60\% & 9.3\%  & 56.8\% & 56.9\% & 84.4\% & 60.4\% & 50.3\% & 29.1\% \\ \hline
	2000 & 68.6\% & 7.2\%  & 62.4\% & 65.3\% & 82.5\% & 68.6\% & 59.8\% & 40.3\% \\ \hline
	3000 & 73.5\% & 6.2\%  & 65.6\% & 70.2\% & 84\% & 73\% & 66.4\% & 44.4\% \\ \hline
	4000 & 83.2\% & 5.2\%  & 78.1\% & 80.6\% & 87.9\% & 83\% & 72.9\% & 43.7\% \\ \hline
	5000 & \textbf{84.5\%} & 5.3\%  & 78.3\% & \textbf{88.9\% }& \textbf{88.5\% }& \textbf{91.8\%} & 78.8\% & 41.6\% \\ \hline
	\end{tabular}\\
	\label{approaches:table1}
\end{table*}

\begin{table*}
	\centering
	\caption{Output of sample accuracy generated through 10-fold cross validation using a training dataset of 5000 tweets.}	
	\begin{tabular}{|l|c|c|c|c|c|c|c|c|}
	\hline
	N-grams & SVM & ME & Tree & Bagged Tree & Boosted Tree & RF & NN & NB \\ \hline \hline
	Uni-gram & 84.5\% & 5.3\%  & 78.3\% & \textbf{88.9\%} & 88.5\% & \textbf{91.8\% }& 78.8\% & 41.6\% \\ \hline
	Bi-gram & 53.7\% & 2.3\%  & 50.9\% & 53.7\% & \textbf{96.5\%} & 52.9\% & 53.5\% & 41.3\% \\ \hline
	Tri-gram & 44.2\% & 37.4\%  & 43.9\% & 44.4\% & \textbf{99.6\%} & 44.3\% & 44.2\% & N/A \\ \hline
	\end{tabular}\\
	\label{approaches:table2}
\end{table*}

\subsection{Modelling}
Eight predictive models are developed using the training data set obtained from machine learning approaches, namely Support Vector Machine (SVM), Na{\"i}ve Bayes (NB), Maximum Entropy (ME), Tree, Bagging, Boosting, Random Forest (RF) and Neural Network (NN). With the exception of Naïve Bayes (NB), which is developed with e1071 package \cite{dimitriadou2006e1071}, all the other classifiers are built with RTextTools \cite{jurka2011rtexttools}, a machine learning library for text classification.

SVM uses kernels to find a hyperplane that divides data into different categories with the maximum margin \cite[chapter 9]{james2013introduction}. In RStudio this model is built with the \textit{train\textunderscore model} function in RTextTools by calling the SVM algorithm from e1071.

NB classifier is a simple probabilistic classifier applied with Bayes Theorem with the assumption that all features are independent to each other \cite{pang2002thumbs}. Although this assumption may not be realistic in real-world situations, studies show that it outperforms other models for certain problems \cite{Pak}. This model is built with the NB function in e1071.

ME is a classification method that generalises logistic regression to a multi-class problem and predicts the probabilities of different possible results. This algorithm is used as a contrast to NB, because of their opposite assumptions about the relationships between features. When the conditional independence assumption of NB fails in a certain case, ME might potentially yield a better result. ME algorithm is called from maxent \cite{jurka2012maxent} via the train model function in RTextTools.

A classification tree is a predictive model using decision tree learning through a process of binary recursive partitioning. This is a popular approach for classification problems in data mining \cite[chapter 8]{james2013introduction}. This is the foundation of other tree-based models that will be discussed later, including bagging, random forest and boosting. Tree algorithm is called from tree \cite{ripley2005classification} via the train model function in RTextTools.

Bagging, also called bootstrap aggregating is an ensemble algorithm used in machine learning to improve the accuracy and stability of an existing model. In this case a bagged classification tree is trained with ipred \cite{peters2009ipred} and RTextTools.

Similar to bagged tree, RF is another ensemble learning algorithm. Instead of fully growing each binary tree, RF controls the number of features to search over in its pursuit of the best split for each tree. Using 200 trees, RF algorithm is called from randomForest \cite{liaw2002classification} via the \textit{train\textunderscore model} function in RTextTools. 

To reduce the bias of a predictive model against incorrectly classified data, boosting algorithms aim to build a strong classifier on a weak one by iteratively adding weights to misclassified data \cite[chapter 8]{james2013introduction}. Logit boosting algorithm from caTools \cite{tuszynski2013package} is employed here to build a boosted tree.

NN performs non-linear statistical modelling and is robust against noise and generalisation \cite{basheer2000artificial}. In this research we apply NN for text categorisation, which is seldom done. NN algorithm is called from nnet \cite{nnet} via the \textit{train\textunderscore model} function in RTextTools.

\subsection{Evaluation}
Table \ref{approaches:table1} compares model performance under 40 conditions generated by different training data sizes of $1000, 2000, \cdots, 5000$ tweets and algorithms using the uni-gram feature. Model performances enhance along with larger training data set, with the only exception of ME. The top four performing models from this evaluation are highlighted and chosen for classifying the data set - the best performing model is RF with nearly 92\% accuracy, followed by Bagged Tree and Boosted Tree which have almost similar performances and the SVM which has almost 85\% accuracy.

As shown in Table \ref{approaches:table2}, model performances are not improved by increasing N-gram features, with the exception of Boosted Tree reaching over 99\% accuracy with tri-grams. The accuracy of ME using tri-gram feature, which is the other exception, and is nearly 20 times higher than using bi-gram. There is an ongoing debate on whether N-gram feature can effectively enhance model performance or not \cite{pang2002thumbs}. On one hand, uni-gram features cover more data where as bi-gram and tri-gram features better capture patterns of expressions. In this research, the uni-gram features are clearly the most effective and it is believed that the choice of N-gram features should depend on the specific corpus. When selecting well-performing models, the two exceptions above are regarded as special cases that cannot represent the general trend. Therefore, the four highlighted best classifiers in Table \ref{approaches:table1}, namely RF, Bagged Tree, Boosted Tree and SVM, are accepted for classifying the test data, the results of which will be compared visually in the web application. 

Although NB is strongly recommended in \cite{wang2012baselines} as a better choice for snippet sentiment tasks than SVM, it fails in this case. Two factors might account for this result: first, the data sets used in other studies are usually a mixture of topics, such as movie reviews; second, the snippet sentiment analysis explores the short version of documents, which contain much less informal languages than tweets. Given the fact that this research focuses on the public sentiment towards one specific topic extracted from tweets, it is possible that so-called strong classifier such as NB could actually fail. NB using tri-gram feature is not available due to the lack of decisive phrases.

\section{Twitter Analysis of Sandy Hook Elementary School Shooting}
\label{casestudy}
In this paper, we analyse the tweet corpus purchased from GNIP surrounding the Sandy Hook elementary school shootings of Friday, 14 December, 2012 at 0940 local (14:40 GMT) in Newtown, Connecticut, USA and visualise the results as a web application which is available at http://www.gunsontwitter.com.

\subsection{Analysing the Tweets}
The acquired Twitter corpus contains 700,437 tweets for a 40 day period from Friday, December 7 2012, 00:00:01 GMT until Tuesday, January 15 2012, 23:59:59 GMT. The relevant tweets were obtained from GNIP by applying a number of filter rules. The rules can incorporate a number of keywords (for example, Sandy Hook, Newtown and guns), phrases (for example, gun violence and firearm magazine), hashtags (for example, \#SandyHook and \#CTshooting) and reply/mention tags (for example, @sandyhook and @SHWeChooseLove) along with a variety of filters such as the country of origin of the tweet (country\_code:us), language (lang:en) used for tweeting and a means to specify that the tweet is or is not a repost (-(is:retweet) as no retweets were selected). 

During pre-processing relevant information such as the tweets, date, time and location were extracted. A golden training data set of 5000 tweets were manually selected and classified on the basis of the following three sentiments towards the event
\begin{enumerate}[leftmargin=0.4cm]
\item Pro-gun: favouring the freedom of gun ownership and against gun control. For example, ``The only thing that stops a bad guy with a gun, is a good guy with a gun'' or ``Gun control isn't gonna help. God bless the 2nd amendment! \#sorrynotsorry''.
\item Anti-gun: favouring gun control and express outrage and sympathy to the event. For example, ``I cannot fathom how one could be so vile that they open fire on children or anyone. We need STRICT gun controls. \#Newtown'' or ``BAN GUNS!!! Let our children be safe''.
\item Neutral:  irrelevant in the context or neither convey a pro-gun or anti-gun sentiment. For example, ``Obama is possibly visiting Newtown'' or ``Not sure if gunshot or firework''.
\end{enumerate}

The four machine learning approaches that demonstrated best performance during evaluation in Section \ref{mlapproaches}, namely RF, Boosted Tree, Bagged Tree and SVM, were used to train models with the training data and then applied to classify the remaining tweets.

\begin{figure}
	\centering
	\includegraphics[width=0.47\textwidth]{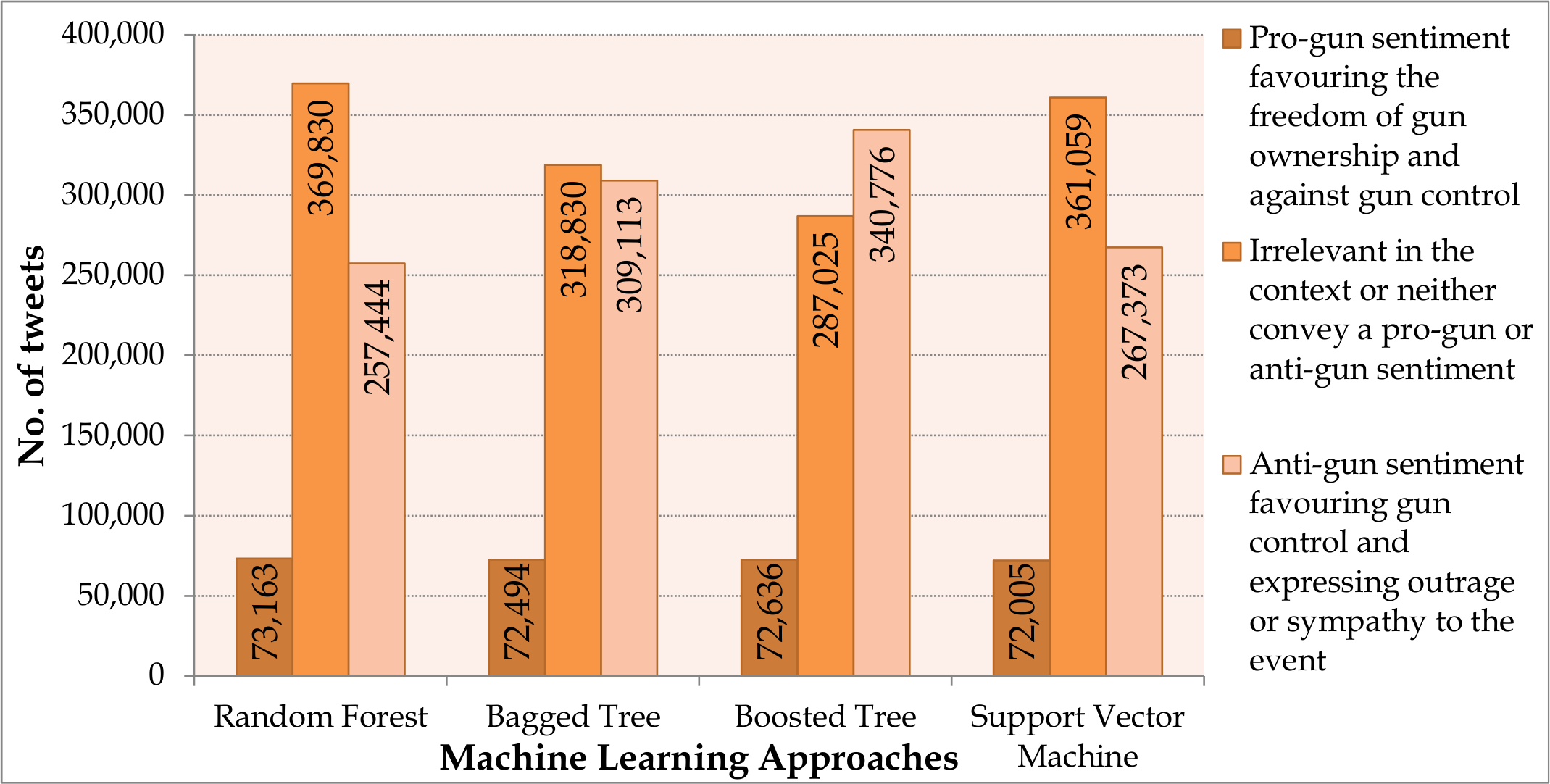}
	\caption{Number of classified tweets from December 7 2012, 00:00:01 GMT to January 15 2012, 23:59:59 GMT}	
	\label{analysis:figure1}
\end{figure}

Figure \ref{analysis:figure1} highlights the numbers of classified tweets using the four selected machine learning approaches. Pro-gun tweets account for about 10\% of the data set regardless of approach chosen. The major difference between the approaches is in the number of neutral and anti-gun tweets. In the evaluation stage, a difference in accuracy by 10\% was observed between the RF and SVM approaches. However, both these approaches assign over half of the tweets into the neutral class. The bagged tree model classifies nearly the same volume of tweets as neutral and anti-gun.

\begin{figure*}
	\centering
	\subfloat[Hashtags]{\includegraphics[width=0.44\textwidth]{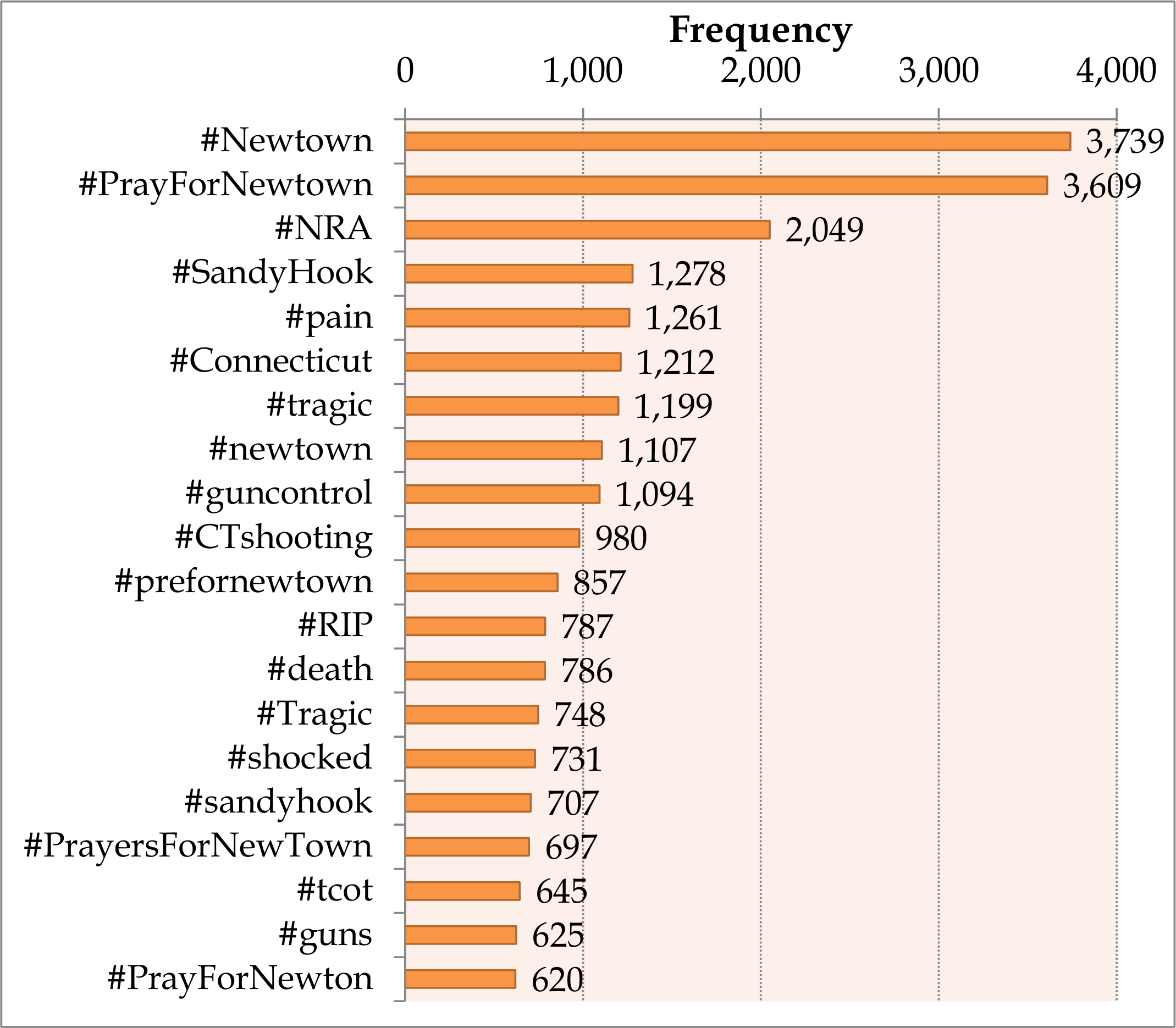}} \qquad
	\subfloat[Reply/Mention Tags]{\includegraphics[width=0.44\textwidth]{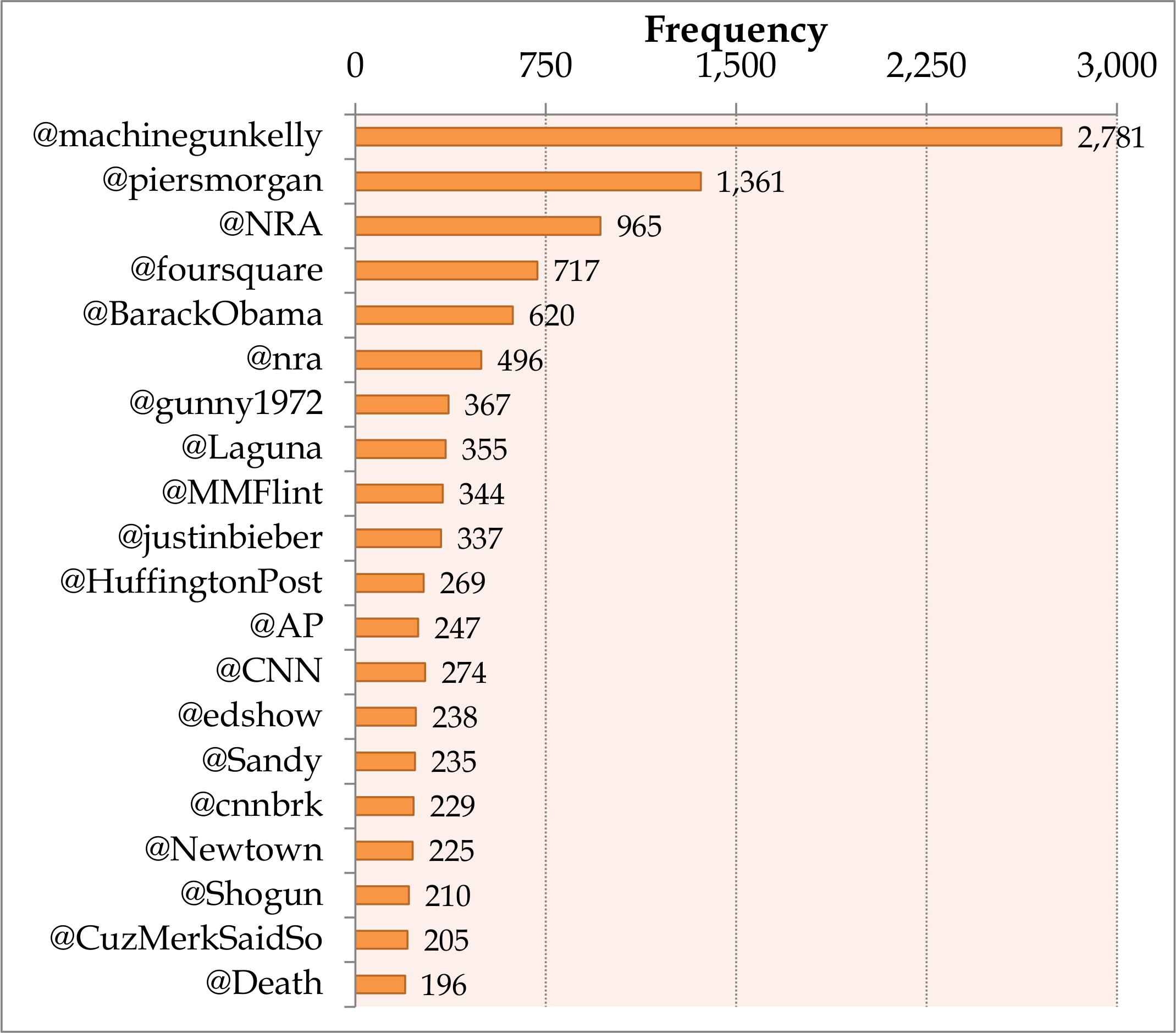}}
	\caption{Top 20 tags frequently used in tweets from December 7 2012, 00:00:01 GMT to January 15 2013, 23:59:59 GMT}		
	\label{analysis:figure3}
\end{figure*}

Figure \ref{analysis:figure3} summarises the most frequently used hashtags and reply/mention tags in the data set. Although some of these tags do not convey obvious sentiments they act as indicators of emotions to some extent. For example \#guncontrol and @BarackObama usually appear in anti-gun tweets while \#NRA is more favoured in tweets that are more pro-gun and advocate gun ownership and gun rights. It would seem that there is more neutral sentiment conveying reply/mention tags (for example @justinbieber or @machinegunkelly). 

\subsection{Interactive Visualisation}
An interactive web application (app), which is available on http://www.gunsontwitter.com, was developed to display multiple dimensions of the data. Data on population\footnote{ https://www.census.gov/popest/data/state/totals/2012/} and percentage of gun ownership (as of 2007)\footnote{ http://usliberals.about.com/od/Election2012Factors/a/Gun-Owners-As-Percentage-Of-Each-States-Population.htm} in each state were incorporated to add value to the visualisation. The following sections demonstrate the three visualisation techniques.



\subsubsection{Motion Chart}
\begin{figure*}
	\centering
	\subfloat[Bubble view]{\includegraphics[width=0.438\textwidth]{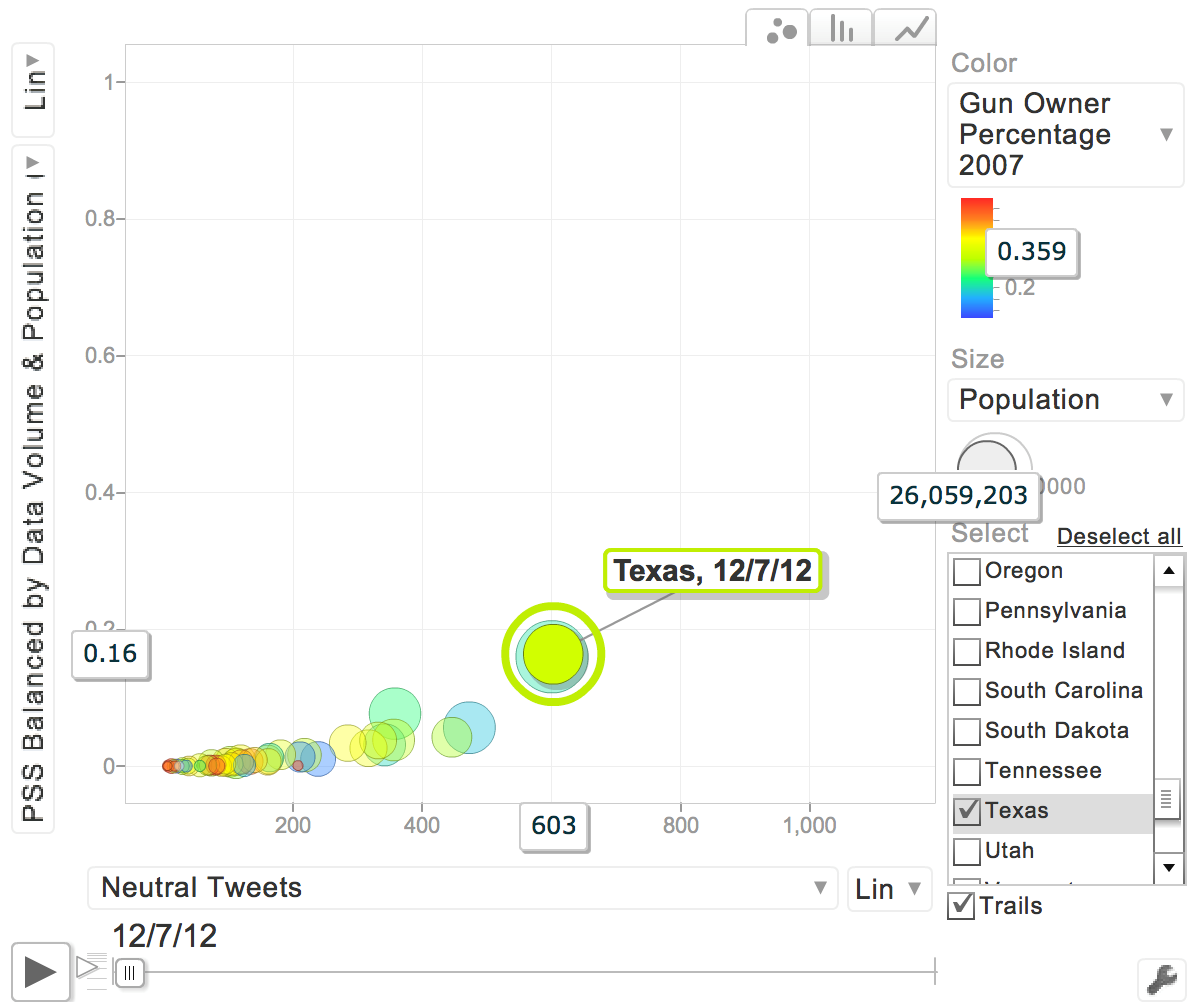}\label{analysis:figure5a}} \qquad
	\subfloat[Trail feature of the bubble view]{\includegraphics[width=0.438\textwidth]{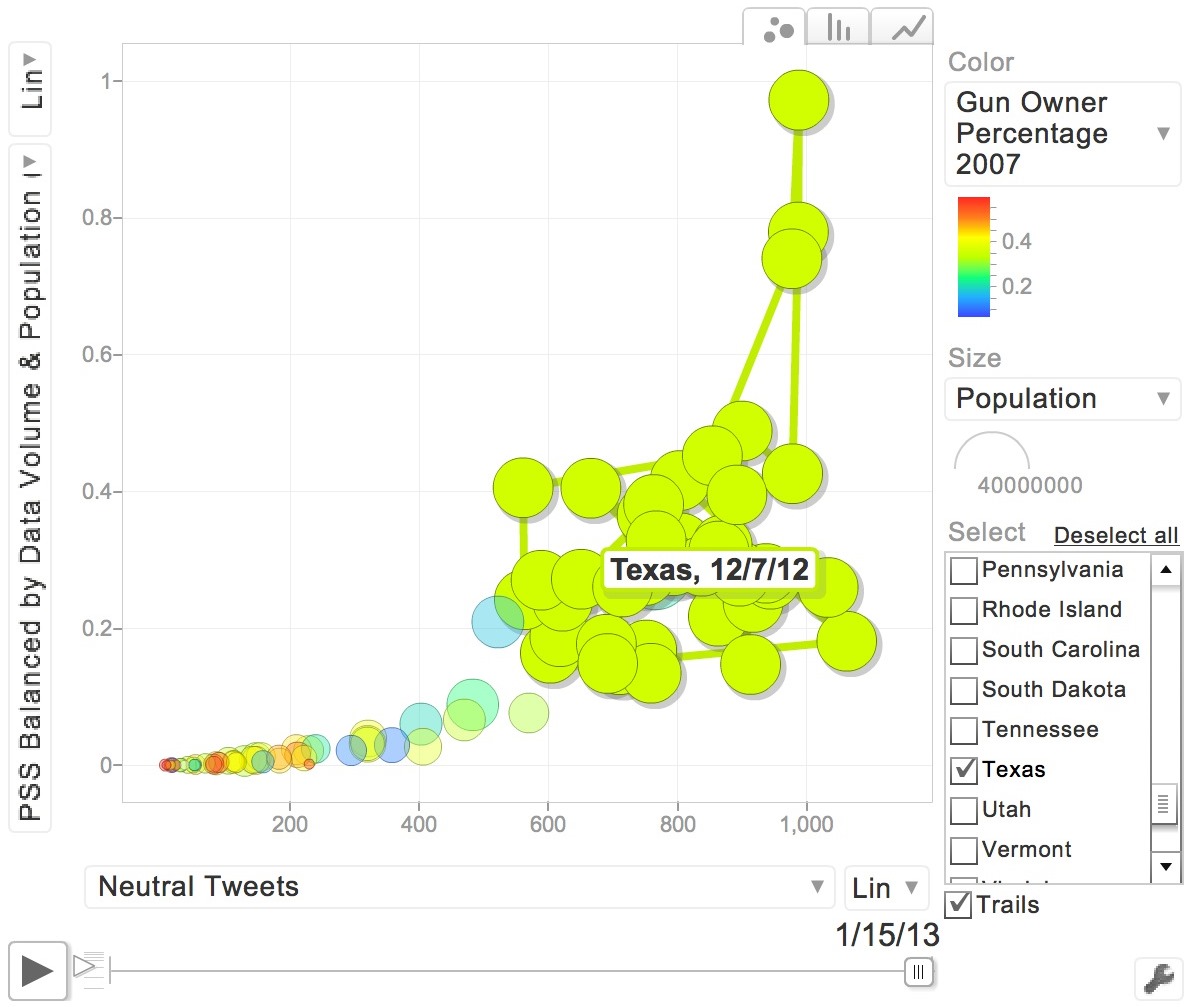}\label{analysis:figure5b}} \\
	\subfloat[Bar chart view]{\includegraphics[width=0.438\textwidth]{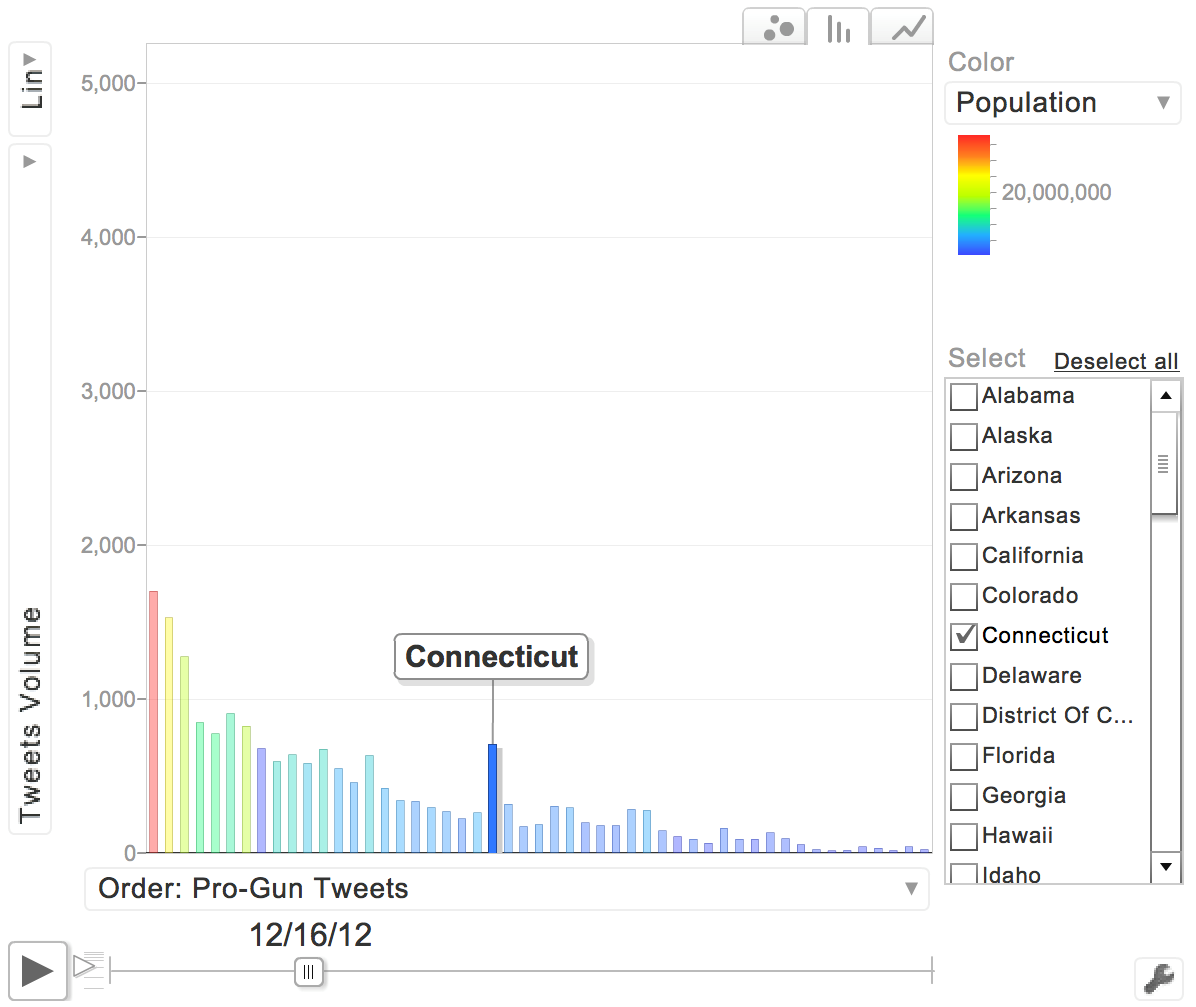}\label{analysis:figure5c}} \qquad
	\subfloat[Line graph view]{\includegraphics[width=0.438\textwidth]{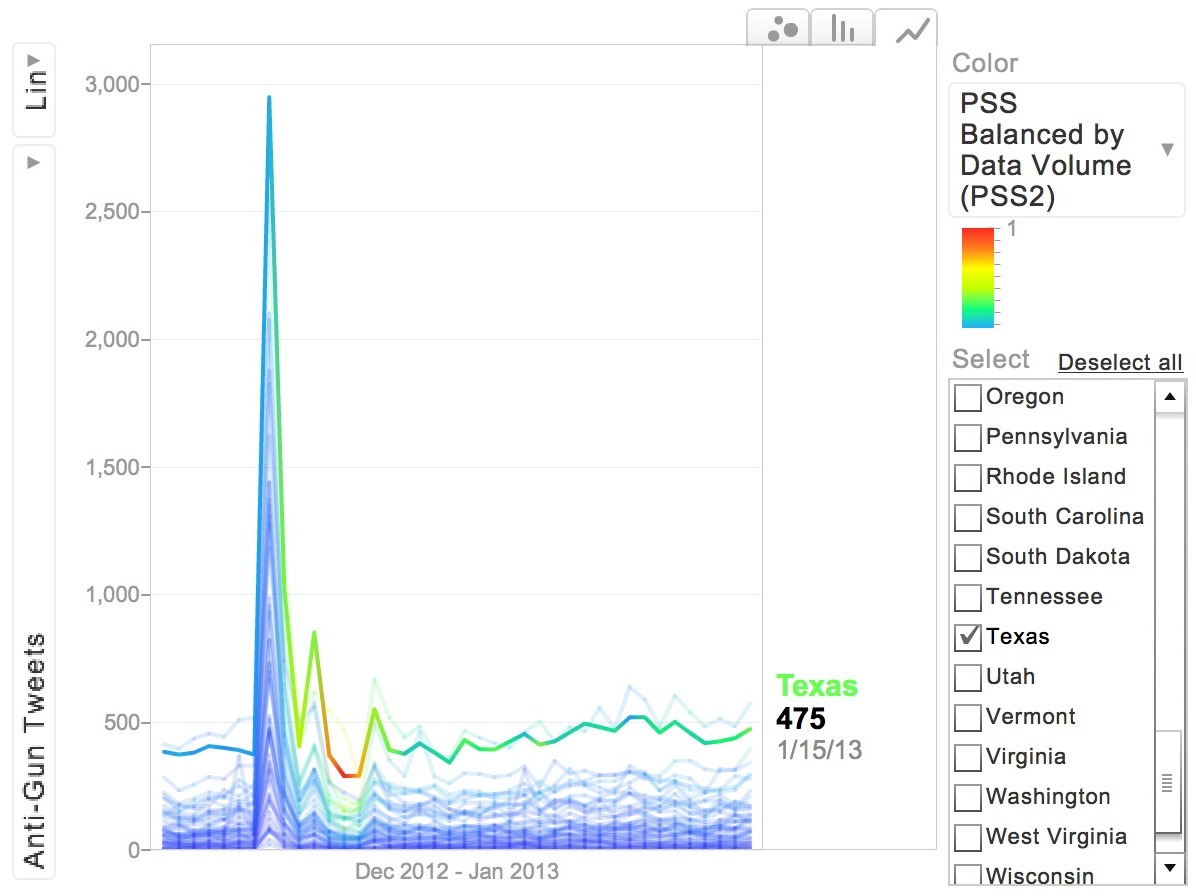}\label{analysis:figure5d}}
	\caption{Motion chart of the interactive web application}		
	\label{analysis:figure5}
\end{figure*}

Figure \ref{analysis:figure5} is an example of the different visualisations possible using motion chart. For these figures the RF approach is selected. Figure \ref{analysis:figure5a} is a bubble view with six variables visualised simultaneously: X-axis for numbers of neutral tweets, Y-axis for public sentiment scores balanced by data volume and population ($PGPSS_3$), bubbles for the states, sizes of bubble for the populations of the states, colours of bubbles for the gun ownership as percentage of population and the slider at the bottom for date. In this chart for December 7 2012 (one week before the shooting event), larger the population of the state more pro-gun is the sentiment for gun use. In contrast, states that have a large volume of registered gun owners show little pro-gun sentiment.

A trail feature can be set by the user for viewing the variation of any variable over a time period. For example, refer Figure \ref{analysis:figure5b}. Texas is selected, and an overview of its variation during the 40-day period is seen on the chart. The highest $PGPSS_3$ appears when the number of neutral tweets increased to its top level.

Figure \ref{analysis:figure5c} shows the embedded bar chart view provided by the motion chart. Five groups of statistics are presented: numbers of pro-gun tweets on X-axis, volumes of tweet on Y-axis, states as bars, populations in the colours of bars and dates on the slider. In this example, the results of tweets on December 16 2012 are presented. The majority of tweets come from states with large population except Connecticut standing out in the middle of X-axis. This is probably because the shooting happened there and people were more concerned.

Figure \ref{analysis:figure5d} is the line graph view of the motion chart. Four variables are available on this view: time series on X-axis, numbers of anti-gun tweets on Y-axis, states shown in lines and public sentiment scores balanced by volumes of data ($PGPSS_2$) represented as the colour of the lines. This graph compares the statistics at the state-level and cannot provide a holistic view at the country-level. Therefore, a different line graph is generated as the second visualisation technique to analyse the results of U.S. at the country-level.

\subsubsection{Line Graph}
\begin{figure*}
	\centering
\subfloat[Daily Analysis]{\includegraphics[width=\textwidth, frame]{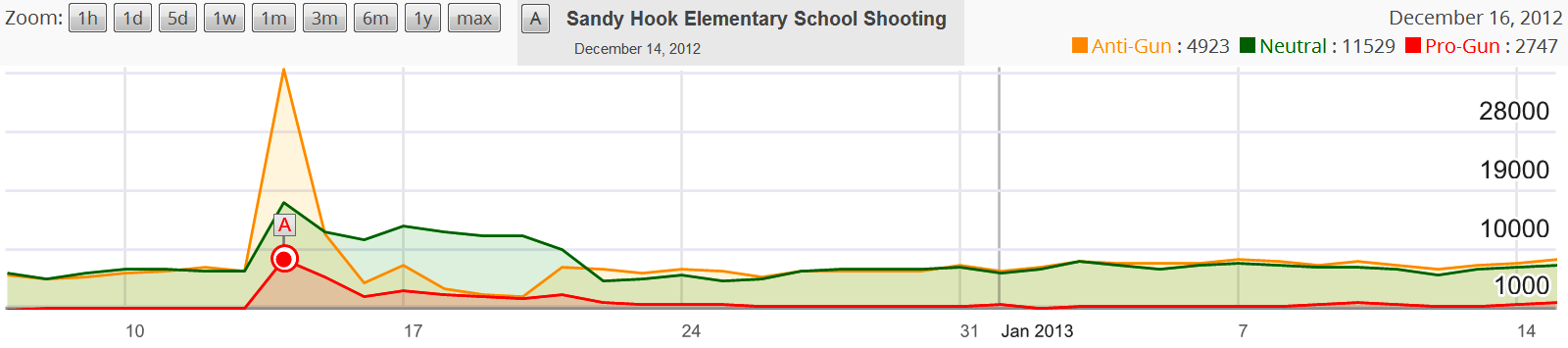}\label{analysis:figure6a}}\\
\subfloat[Hourly Analysis]{\includegraphics[width=\textwidth, frame]{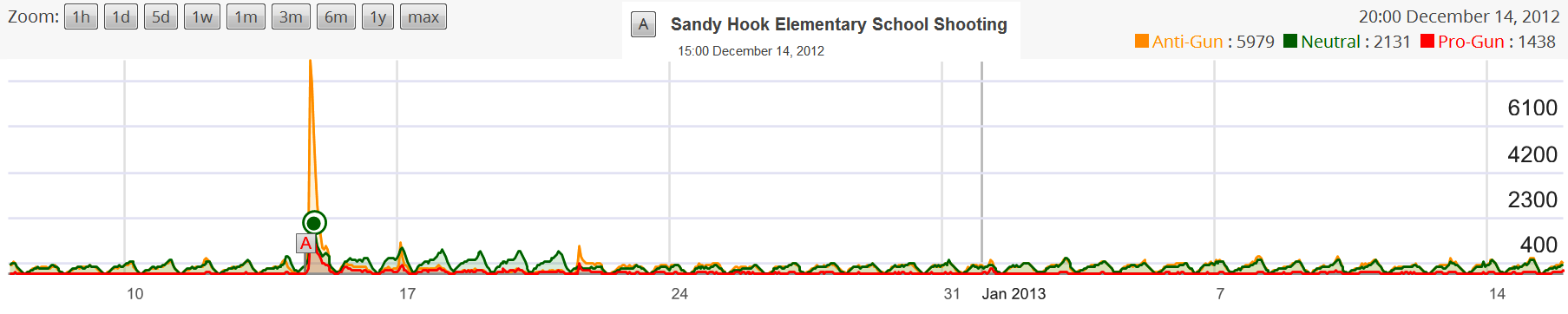}\label{analysis:figure6b}}
\caption{Line graph of the interactive web application}
	\label{analysis:figure6}
\end{figure*}

Figure \ref{analysis:figure6} presents the visualisation using line graph. These results are presented for the Bagged Tree model. The numbers of tweets in the three sentiment classes can be explored either on a daily (Figure \ref{analysis:figure6a}) or an hourly (Figure \ref{analysis:figure6b}) basis. Zooming options are realised through a series of buttons at the top left corner, from one hour to the maximum time period. The time frame at the bottom is designed to customise the view a certain period. By default the line graph displays the results for the 40-day period. These graphs highlight that during the one-week period past the shooting, the number of tweets increase considerably regardless of the sentiment (pro-gun or anti-gun) with peaks observed on December 14 2012. An interesting fact is that although anti-gun sentiment was normally stronger than the pro-gun sentiment, on December 19 2012 the anti-gun sentiment is as low as the pro-gun sentiment. This five-day period (from December 14 to 19) could be important for using public opinion as an instrument for facilitating policy making.

\subsubsection{Geographic Map}
\begin{figure*}
	\centering
	\subfloat[$PGPSS_1$]{\includegraphics[width=0.33\textwidth]{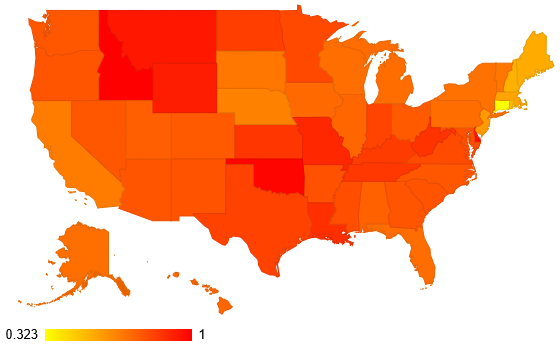}}
	\subfloat[$PGPSS_2$]{\includegraphics[width=0.33\textwidth]{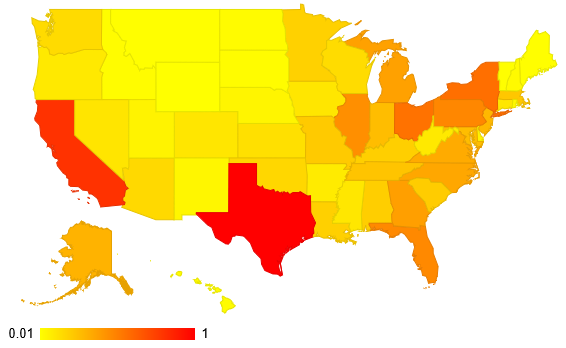}} \hfill
	\subfloat[$PGPSS_3$]{\includegraphics[width=0.33\textwidth]{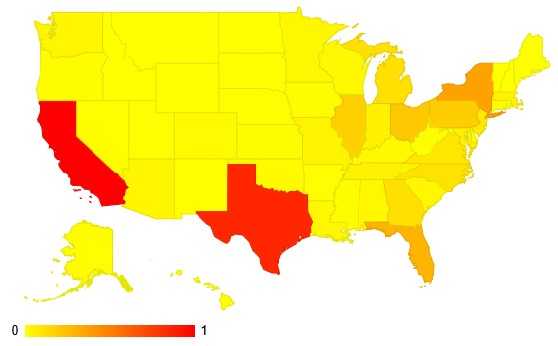}\label{analysis:figure7a}} \hfill
	\caption{Geographic map with customised PGPSS for December 7 2012 to January 15 2013}		
	\label{analysis:figure7}
\end{figure*}

\begin{figure*}
	\centering
	\subfloat[$PGPSS_1$]{\includegraphics[width=0.33\textwidth]{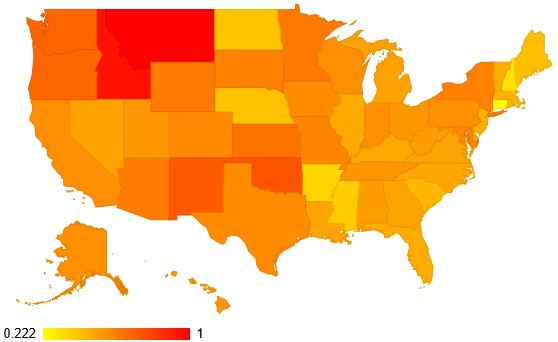}}
	\subfloat[$PGPSS_2$]{\includegraphics[width=0.33\textwidth]{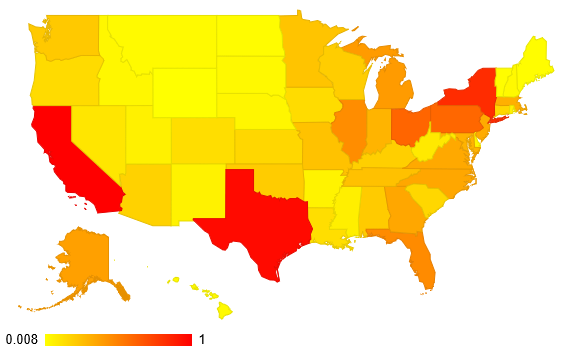}} \hfill
	\subfloat[$PGPSS_3$]{\includegraphics[width=0.33\textwidth]{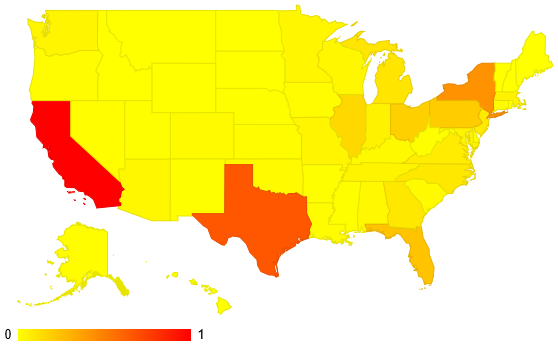}\label{analysis:figure8a}} \hfill
	\caption{Geographic map with customised PGPSS for December 13 2012 to December 15 2012}		
	\label{analysis:figure8}
\end{figure*}

The coordinates of each tweet (latitude and longitude) were used to categorise the tweets from each state. This was facilitated using four R packages, namely regos \cite{rundel2013package}, sp \cite{sp}, maps \cite{maps} and maptools \cite{maptools}. The coordinate information was used to generate a geographic visualisation as shown in Figure \ref{analysis:figure7} and Figure \ref{analysis:figure8}.

Figure \ref{analysis:figure7} presents three maps based on $PGPSS_1$, $PGPSS_2$ and $PGPSS_3$ for the 40-day period. $PGPSS_3$ is the best estimation since it is corrected for both volume of data and population. $PGPSS_1$ may not be a good indicator since it only takes the ratio of pro-gun and anti-gun tweets into account. In addition, the percentage of gun ownership appears in a floating box when the cursor hovers over a state. By default the geographic map displays $PGPSS_3$ over the whole 40-day period until the user chooses dates of interests through the date picker. For example, refer Figure \ref{analysis:figure8}. By comparing Figure \ref{analysis:figure7a} and Figure \ref{analysis:figure8a}, the three states that have strong pro-gun sentiment are California, Texas and New York.

\section{Discussion}
\label{discussion}
The Sandy Hook school massacre was a deeply distressing and tragic event. It raised issues related to gun ownership and mental health. The mental health issue is not addressed by our data but we note that others who have carefully considered the issue \cite{gun-1} conclude that ``notions of mental illness that emerge in relation to mass shootings frequently reflect larger cultural stereotypes and anxieties...''. The focus thus logically returns to gun ownership and access. High profile events typically evoke a huge and often opinionated social media response \cite{gun-2, gun-3, gun-4}. In instigating the analysis our prior assumption was that there would be a peak of anti-gun sentiment expressed in the hours immediately following the shooting. This did indeed turn out to be the case. We also assumed that after a period of time there may be a pro-gun response; that this response may come in reply to calls for greater gun control and would be seen by those involved as a reassertion of their second amendment rights. Our assumption was that this response may have occurred after about 7 to 10 days and if so we were interested in delineating the period between the anti-gun and pro-gun spikes as this perhaps defined a period during which preventative political action could be initiated.

Surprisingly, at least to us, the pro-gun spike in sentiment was not delayed but actually occurred on the day of the killings. We do not assume that this was a heartless response to a truly shocking and distressing event. It is quite possible that some of those expressing pro-gun sentiments were unaware that there had been fatalities at the time they tweeted. For good operational and practical reasons the fact that there had been so many deaths did not start to appear in the media until quite some hours after the event. 

It is also interesting to note that whilst anti-gun sentiment peaked and then quickly fell to pre-shooting levels, pro-gun sentiment ran at an elevated level for some time. Again we think this probably reflects a genuine and deeply ingrained difference in opinion between individuals about the role of guns in causing or preventing such mass shooting events.

Indeed we think it is likely that pro-gun tweeters on the day of the killings did so because they genuinely believe that the event could have been avoided had members of staff been armed. A full discussion of the evidential basis of any such presumed belief or indeed of its more general moral, social and educational consequences is beyond the scope of this paper. State level analysis showed considerable variation in response. The highest gun owning states were not the most pro-gun in this situation. This may be because reasons for gun ownership vary considerably. Some traditional hunting states with high level of gun ownership demonstrated relatively low levels of pro-gun sentiment. California, Texas and New York demonstrated high levels of pro-gun sentiment. A full exploration of this interesting observation is beyond the scope of the data at the heart of this paper. 

\section{Related Work}
\label{relatedwork}
Sentiment analysis of social media is an important avenue of research in the area of natural language processing. Previous research includes document-level analysis of product reviews \cite{pang2008opinion} to term-level inspection on the polarity of words \cite{esuli2006sentiwordnet} \cite{mohammad2009generating}. Since tweets are character limited sentence-level classifications are frequently adopted to extract public sentiment. Classifying tweets is challenging due to the unique nature of micro-blogging with its frequent use of informal and colloquial language including slang and emoticons. Although there are different approaches for classifying tweets, there is no consensus on the best solution. In natural language processing of tweets, engineering linguistic features and automating text categorisation are two important tasks.

\subsection{Linguistic Features}
Twitter sentiment has been explored from the traditional linguistic perspectives and extracting N-gram features is a baseline method among researchers \cite{kouloumpis2011twitter}. Uni-gram is the most common feature used and is reported to result in good predictive models with accuracy around 70\%-80\% \cite{sakaki2010earthquake} \cite{saif2011semantic}. Bi-gram and tri-gram features are claimed to be effective in improving model performances. However, research suggests that uni-grams are better when building classifiers~\cite{pang2002thumbs}.

Other features such as Part-Of-Speech (POS) tags \cite{Go} and the position of words in text \cite{pang2002thumbs} are strong indicators of emotional text \cite{Pak}. However, experiments show that POS tags may not be useful for sentiment analysis of micro-blogs \cite{kouloumpis2011twitter}. Hash-tags and emoticons are typical contents in Twitter that indicate latent sentiment which are made use of in the process of automatically generating training data~\cite{bifet2010sentiment} \cite{davidov2010enhanced}.

\subsection{Automatic Text Categorisation}
Algorithms for automated classification include dictionary-based and machine learning approaches. Dictionary-based approaches \cite{tumasjan2010election} \cite{asur2010predicting} compare tweets against lexicons of existing dictionaries. By counting frequencies of the words matched with the chosen dictionary, a tweet is assigned a sentiment based on the prevalent emotion in the text. This approach is disadvantageous in that the acquired lexicon is domain independent and therefore it is impossible for a dictionary-based algorithm to capture different contexts. Another drawback of this approach is that it has to ignore the informal language used in tweets, which could probably stand for strong emotions. For example, words with repeated letters such as `noooooooo' convey a stronger emotion than `no'. To overcome these disadvantages, machine learning based approaches provide the possibility of enhanced utility.

Machine learning approaches can be incorporated in building classifiers and can produce accuracy comparable to human experts \cite{sebastiani2002machine}. However, the best classifier depends on the specific case study it is employed for; SVM and NB are popularly used owing to their average accuracy of above 80\% regardless of features selected \cite{Go}. Other algorithms such as Conditional Random Field (CRF) and Maximum Entropy (ME) are found to be less effective for tweets analysis \cite{Pak} \cite{Go}.

In addition to the aforementioned standard machine learning techniques, adaptive classifiers have been built to improve performance. For example, the NBSVM, an SVM classifier trained with NB features, has better performance than SVM \cite{wang2012baselines}. Similarly, an interpolation method \cite{saif2011semantic} can be employed to modify the NB classifier with a generative model of words given semantic concepts. Hybrid classification has been proposed as an effective technique to combine a rule-based classification and machine learning approach \cite{prabowo2009sentiment}.

\subsection{Challenges}
Although functional techniques have been developed in the area of using machine learning for sentiment analysis, there continues to be three challenges that we seek to address in this paper. Firstly, there is a lack of evaluation of various machine learning approaches in the same Twitter analysis context. For example, the so-called best classifiers, which were trained for categorising positive and negative text, demonstrated poor performance with three classes (positive, negative and neutral) \cite{Go}. Meanwhile, approaches such as Neural Network (NN) and Boosting in sentiment analysis are scarcely investigated. Hence, we believe it is essential to evaluate different approaches in the same context in order to identify the best approach in any given context.

Secondly, an important social issue, namely gun violence, has not been investigated for understanding public sentiment. Twitter analysis has been employed for understanding complex phenomena in the areas of politics \cite{tumasjan2010election}, disaster management \cite{sakaki2010earthquake} and public health \cite{paul2011you}. Gun violence is an important and challenging topic in public health and understanding the temporal and geographic patterning of pro-gun and anti-gun sentiments is important in terms of framing and timing political initiatives aimed at reducing gun related deaths. This is particularly true in terms of mass shootings and in this paper, we seek to use machine learning based Twitter analysis to explore pro-gun and anti-gun sentiments in relation to the Sandy Hook School shootings.

Thirdly the use of machine learning techniques for the analysis of large data sets is in general still confined to the academic discipline of Computer Science (CS). In seeking to bridge the disciplines of Computer Science and Public Health in producing this paper we are seeking to further the prospects of such trans-disciplinary research by promoting the benefits that machine learning can offer. 

Hence, in this study, the above three challenges related to the evaluation of machine learning techniques, the application of Twitter analysis to gun violence and the accessibility of machine learning techniques for non-CS experts are addressed through the Sandy Hook Elementary School shootings case study. Tweets expressed public sentiment in terms of pro-gun or anti-gun attitudes were examined over forty days. These sentiments were extracted from tweets using machine learning techniques and visualised.

\section{Conclusions}
\label{conclusions}
The public health implications are firstly methodological. The disciplines of Computer Science and Public Health are brought together on the important and emotive issue of reactions to a mass school shooting which is a useful exercise that yields multidisciplinary dividends. 

Furthermore in this paper, we have evaluated a number of machine learning approaches and identified those most suitable to classifying public sentiment towards gun violence in light of the Sandy Hook school shooting. We have shown that it is possible to analyse a large body of social media data using machine learning in a reliable and replicable way by employing a methodology to collect, train and classify tweets. Public sentiment is captured for different time scales and by state can be interactively visualised on an online tool available at http://www.gunsontwitter.com.

Sentiment analysis shows a peak of anti-gun feeling on the day of the Sandy Hook school shooting which quickly falls to pre-event levels. More surprisingly the analysis shows a peak of pro-gun sentiment on the day of the shooting that is sustained at an elevated level for a number of days. These findings suggest that a discernible minority of the US population see high levels of public gun ownership as part of the solution rather than part of problem. This is in spite of the objective evidence that the US with its generally liberal gun ownership rules suffers a uniquely high incidence of school shootings \cite{gun-5} and that high level of public firearms ownership increase firearm related deaths both of the public \cite{gun-6} and of law enforcement officers \cite{gun-7}. 

More work is required to understand the complexities of public opinion on this important issue but social media analysis appears to have a useful contribution to make. In the future, we aim to apply machine learning techniques to additional gun violence related events. This will enhance our knowledge of public reaction and delineate a time period during which preventative political action can be instigated. 


\bibliographystyle{ieeetr}
\bibliography{references}

\end{document}